\documentclass[12pt]{article}
\usepackage{axodraw}
\title{Computational Theory of Biological Function I --- The Kinematics of Molecular Trees}

\author{Ron Maimon}

\begin{document}
\maketitle
\abstract{ This series presents an approach to mathematical biology which makes precise
the function of biological molecules. Because biological systems compute, the theory is
a general purpose computer language. I build a language for efficiently representing the
function of protein-like molecules in a cell. The first paper only presents the kinematic
part of the formalism, but this is already useful for representing large-scale protein
networks. The full formalism allows us to investigate the properties of protein interaction
models, ultimately yielding an estimate of the random-access memory of the proteins, a
measure of their capacity for computation.
}

\section{On Biological Function}

We can understand the struggle for existence which characterizes the biological world
as a competition for computational resources. The computer is the organism, and the
program is its behavior. The computing on the cellular level is done by molecules and
molecular complexes, which control the transformation of the information encoded in
their configurations. Although the computation is internal, the resources to continue
it come from outside.
 
If we view life in this way, the goal of an organism is to reorganize as large a part
of the physical world as possible to perform its computation. Replication is one way of
arranging this, but so is growth and uptake of nutrients--- any process which lets
the program of the organism expand. Because the total available memory and processing
is limited, organisms must compete with one another.

A mathematical theory of biological function is a precise description of the internal
program of an organism. The goal of is to describe it abstractly, in a formal language
which describes the data and algorithm, not the physical motions of the parts. Since
the smallest computing objects in a cell are molecules, the first practical objective
is to find a description of molecular interactions.

\section{A Physical Description Is Too Large}

We want to describe the function of biological molecules, so we start with the a physical
description of their state. Biology is classical, so this is ultimately just the position
and velocity of all the atoms. This data is not only enough to predict the biological behavior,
it is clearly much too large.

We might try to find a smaller collection of variables which can still predict the motion,
at least statistically. For example, instead of describing the position of every atom, we
describe the relative angle between amino acids. We might then describe the statistics of
the shape of a molecule by writing equations of motion for the evolution of the angles. But
no matter how we try to truncate the physics, we always end up with many more variables
than a biological experiment would suggest are necessary. If we want to understand biology,
we must understand why.

The reason is that different bits of information in a physical description have vastly different
computational roles, and most of these bits can't compute. Some randomize effectively instantly,
like those that describe the position and orientation of all the free water molecules. These
bits cannot be used to store information. Others are locked up in integrable motion, and
although these bits can store data, they can't manipulate it in a complex enough way.

To illustrate, imagine you are given complete control of the atmosphere of the earth. You set
up enormous fans and arrange the wind to blow any way that you like, and you encode some data
in the velocity. The fans are taken away, you come back a few months later and try to recover
the information you encoded. You fail, because all the information has been erased by the chaos.

On the other hand, if you are given complete control of a regular system, like the moons of
Jupiter, you can arrange the radii and phase of the orbits to store data forever. But if you
now try to find the output of a general program using the moons as your computer, you will fail.
Because the system is regular, the behavior after an arbitrarily long time can be predicted by a
computation of almost fixed size. We transforms to action-angle variables, then time evolve the
system in these coordinates, and transform back. The behavior of a general algorithm at time $t$
can only be predicted by a computation which takes time $t$, so the moons cannot execute an
arbitrary algorithm. A regular system cannot compute.

But neither can a strictly chaotic system, because the chaotic system has no memory. So to
describe a natural computation, we must ignore any behavior which is chaotic or regular.
The biology is whatever is left.

\section{A Theory Is A Language}

The {\em biological state} of a cell is the smallest collection of data structures which we can use
to predict its behavior. The biological state must have an effectively infinite memory capacity, like
a very large regular system, and a dynamics which cannot be predicted except by simulation. The bits
must be neither regular nor chaotic, but computational.


Since every biological system always includes chaotic subsystems, the computation has access to a
random number generator. When we claim that biological systems cannot be predicted, we are making
a statement about computational undecidability. But the undecidability of random computation
is not an immediate consequence of the undecidability of deterministic computation. In the appendix,
I will show that random computation is undecidable too, although with different time bounds. We can
conclude that the behavior of a general biological system is unpredictable in the strongest possible
way.

If we can't predict the behavior, what kind of theory are we supposed to build?

The answer is clear if we consider program size. Although we cannot predict the large-scale
and long time behavior of a computation, we can find effective description for the parts. Just
as we ignore non-computing physical bits to arrive at the biological state, we can also try to
minimize the number of functional bits once we know the algorithm. The goal of theory is to
approach the smallest possible computational description.

A mathematical theory of biological function then is a language, one that describes the computation
of the biological system as efficiently as possible. Because the system is complex enough to embed a
general purpose computer, we need a general purpose computer language.

\section{Three Axioms For a Language of Types}

I divide the molecules in the cell into two types, {\em protein-like} and {\em RNA-like}. The
RNA-like molecules are those that carry sequence--- the nucleic acids. If we want to describe the
dynamics of these molecules, we need to know what sequence they carry. The protein-like molecules
include sugars, lipids, and proteins, whose interactions do not depend on the precise sequence, but
on the conformation and binding of domains.

Protein-like molecules come in distinct types, and two molecules of the same type potentially
carry the same information. So I only need to describe all the configurations of each of the
different types, and from that I should be able to reconstruct the configurations of the
individual molecules.

There are certain natural properties which a language describing molecular types should have, and I
formalize these in the following axioms:\\
\\
1. I should be able to add a new type to the description whose molecules have only one configuration
and aren't bound to anything. A molecule of this sort is called an {\em atom}.\\
\\
2. If I can describe two types, {\tt A} and {\tt B}, I should be able to add {\tt L=\{ A B\} },
a type called the {\em likebox} of {\tt A} and {\tt B} which describes a molecule which
can be either {\tt A} or {\tt B}.  A molecule of type {\tt L} carries one bit $i$ as part of
its state, and if $i=0$, a state of {\tt A}, and if $i=1$ a state of {\tt B}.\\
\\
3. If I have two types, A and B, I should be able to declare that they are {\em linked}, written
{\tt A-B}. This means that whenever a molecule of type {\tt A} is found, it is always attached to
a molecule of type {\tt B}.\\
\\
These axioms, appropriately interpreted, determine a unique minimal formalism which is apparently
expressive enough to efficiently describe the kinematics of the protein-like molecules in a cell.

I haven't defined the axioms precisely, because the most constructive definition uses the linkbox
normal form, and this is hard to motivate immediately. Until I do this, the discussion will be
intuitive.

Yet even on the intuitive level, the last axiom is ambiguous. Suppose I create three atoms {\tt a}
{\tt b} and {\tt c}, and declare that they are linked in the following pattern {\tt a-b},{\tt b-c},
{\tt c-a}. Every time I have a molecule of type {\tt a}, it must be linked to a molecule of type
{\tt b}, which in turn is linked to a {\tt c}, which is linked to an {\tt a}. But is this {\tt a}
a new {\tt a} or the same {\tt a} that started the loop?  If it's the same {\tt a}, then the
linkage only describes the binding of three molecules. If it is a new {\tt a}, the molecules
form an infinite chain {\tt ...-a-b-c-a-b-c-...}.

I will always resolve the ambiguity in favor of infinite chains. Whenever a sequence of linkages
closes a loop, an explorer traversing the loop comes to a different molecule. This means that the
molecules can never bind in a circle, so the formalism will only describe molecular trees. This is
not as much of a restriction as it seems, because any circle of binding can be unwrapped to form a
chain, in the same way that a topological space can be unwrapped along its cycles into its simply
connected cover. By adding some group theory, we will be able to describe the circles too.

We don't need three types to bind to close a loop. If I link two atoms {\tt a} and {\tt b} twice 
{\tt a} is bound to {\tt b} which is bound to another {\tt a} in an infinite chain. This suggests
the right way to interpret self-binding--- {\tt a-a} makes an infinite polymer.

\section{Examples Building Up To The General Case}

Let us form atoms {\tt a},{\tt b},{\tt c},{\tt d} and place them in a likebox ${\tt L_1}${\tt=\{a b c d\}}.
Let us also form atoms {\tt e f g} and place these in a likebox ${\tt L_2}${\tt =\{e f g\} }. I haven't
defined how to form a likebox with more than two elements yet, but there are several obvious ways. I
can build a four element likebox from pairs by making ${\tt \{ \{ a_1 a_2\} \{a_3 a_4\} \} }$, and
nesting further, I can build any power of two. Combining powers of two, I can form any number,
and the number of bits that describe the state of the likebox only grows as the logarithm of its
size.

So I can equally well describe the state of an $n$ element likebox using an integer {\em index} $i$
between $0$ and $n-1$, followed by the state of the i'th element. In particular, a molecule of type
${\tt L_1}$ has four possible states, while a molecule of type ${\tt L_2}$ has three.

Now I will link ${\tt L_1}$ and ${\tt L_2}$. This forms a {\em linkbox} ${\tt N=L_1-L_2}$, which is
the type of the molecule of ${\tt L_1}$ joined to ${\tt L_2}$. Since ${\tt L_1}$ had four states and
${\tt L_2}$ three, a molecule of type {\tt N} has twelve possible states. These are described by two
integers, one for the index of ${\tt L_1}$ and one for the index of ${\tt L_2}$. I will write the
twelve states of {\tt N} as a list of the indices: {\tt N<0,0> N<0,1> N<0,2> N<1,0> ... N<3,2> }.

If I form a linkbox from $k$ likeboxes with $n_1,n_2,...,n_k$ new atoms, it will have $n_1 n_2 ... n_k$
different states, and these are specified by giving $k$ successive indices, each in the appropriate range.

But in order to form a linkbox from $k$ likeboxes, I have to specify how to link more than two elements. 
I do this by linking all the elements successively in a line. I should not close a loop, or I would
get an infinite chain. 

For the next example, I create atoms {\tt b'} and {\tt b''} and link them. Next, I form two likeboxes
${\tt L_1}${\tt=\{a b'\}} and ${\tt L_2}${\tt =\{c b''\} },and two more likeboxes ${\tt L_3}${\tt =\{p q\}}
and ${\tt L_4}${\tt =\{r s\} } where {\tt p q r s} are new atoms.  Finally, I link the likeboxes to form
${\tt N_1=L_1-L_3}$ and ${\tt N_2=L_2 -L_4}$. What are the configurations?

${\tt N_1}$ can be in two possible states in the first position, either {\tt a} or {\tt b'}, and in two possible
states in the second position {\tt p} or {\tt q}. Likewise, ${\tt N_2}$ has two possible states in the first
position, which are either {\tt c} or {\tt b''} and two in the second position {\tt r} and {\tt s}. If a molecule
of type ${\tt N_1}$ has a {\tt b'}, it is necessarily linked to a {\tt b''}, which itself is part of ${\tt N_2}$,
and therefore links to one of {\tt r} or {\tt s}. So we have described two molecules which can bind each other, and
each carries two other independent configurations.

I will write the four unbound molecules in this way\\
${\tt N_1}${\tt <0,0>}\hskip 50pt ${\tt N_1}${\tt <0,1>}\\
${\tt N_2}${\tt <0,0>}\hskip 50pt ${\tt N_2}${\tt <0,1>}\\
and the four bound complexes as\\
${\tt N_1}${\tt <1-}${\tt N_2}${\tt <1,0>,0>}\hskip 50pt ${\tt N_1}${\tt <1-}${\tt N_2}${\tt <1,1>,0>}\\
${\tt N_1}${\tt <1-}${\tt N_2}${\tt <1,0>,1>}\hskip 50pt ${\tt N_1}${\tt <1-}${\tt N_2}${\tt <1,1>,1>}

I will explain the notation fully in the next section. For now, it's enough to know that the indices resolve
which likebox configuration is carried by each molecule, and the hyphen resolves any linkbox linked to the
current one. This method of writing down the configurations doesn't generalize to arbitrary linkboxes and
likeboxes, but it works for any normal form.

Now suppose I construct atoms {\tt a' a''} and link them. I will now construct a linkbox
{\tt D=\{p a' a''\}-\{r s\} }.  The new thing here is that the first likebox contains
two objects which are themselves linked. What does this represent?

If {\tt D} has a {\tt p} in the first position, there are two possible states, depending on whether the state in the
second position is {\tt r} or {\tt s}. If {\tt D} contains the state {\tt a'}, however, it must be attached to an {\tt a''},
which is necessarily part of another {\tt D}. This second {\tt D} has {\tt a'} in the first position, but it can
have either {\tt r} or {\tt s} in the second. So this describes a molecule which can dimerize, and independently
of the dimerization has two conformations it can assume. There are six different configurations, one of which is
{\tt D<1-D<2,0>,1>}, which describes a dimer with one of the components carrying an {\tt r}, and the other carrying
an {\tt s}. Note that this is a different dimer than {\tt D<1-D<2,1>,0>}.

If I construct a likebox of $k$ linked atoms and place it in a linkbox, the linkbox describes a molecule which
can form a k-mer.

Let me again construct two linked atoms {\tt a'-a''}, and from them form the linkbox\ \ {\tt A= \{p a'\}-\{q a''\}}.
This linkbox can also bind to itself, but now the self binding is through atoms in different likeboxes. A molecule
of type {\tt A} can have a {\tt p} or an {\tt a'} in the first position, and if it contains an {\tt a'}, it is
linked to another molecule of the same type with an {\tt a''} in the second position. We know that this second
molecule has an {\tt a''} in the second position, but we don't know if it has a {\tt p} or an {\tt a'} in the first
position, and if it has an {\tt a'}, it is linked to a third molecule. This third molecule can bind a fourth,
and so on.  So this example is a molecule which can bind into polymers of arbitrary length.  One of the polymers 
is described by the string {\tt A<0,1-A<1,1-A<1,0>>>}.

If I construct a linkbox of $k$ likeboxes, each containing one of a linked chain of atoms, it describes a
molecule which polymerizes into a Bethe-lattice with branching factor $r=k(k-1)$. The number of molecules
$n$ bonds away from any initial one is $r(r-1)^{n-1}$.

For the last example, construct atoms {\tt b' b'' b''' b'''' } and link them in a sequence {\tt b'-b''},{\tt b''-b'''}
,{\tt b'''-b''''}, abbreviated as {\tt b'-b''-b'''-b''''}. Next construct linked atoms  {\tt b'-b''}, and unlinked atoms
{\tt c d e }.  

Form three likeboxes ${\tt L_1}$ {\tt =\{c a' b'\}}, ${\tt L_2}${\tt = \{d a''\}}, and
${\tt  L_3}$ {\tt =\{e  b'' b''' b''''\}} and link ${\tt L_1}$ and ${\tt L_2}$ to form a linkbox
${\tt N_1= L_1-L_2}$. I will consider ${\tt L_3}$ itself to be a linkbox, although it is not linked
to anything else. The linkbox ${\tt N_2=L_3}$ contains exactly one likebox.

The configurations of ${\tt N_1}$ and ${N_2}$ can be understood from the discussion above. ${\tt N_1}$
polymerizes, unless it attaches through {\tt b'} to a trimer of ${\tt N_2}$ molecules. This
configuration describes two molecules which can form a three-pronged trident, which can grow arbitrarily
far at any of the ends.

So it is clear that the three axioms can produce complicated enough patterns of binding. In the next
section, I will systematically study the types that occur.

\section{Linkbox Normal Form}

All the examples were built by following this recipe:\\
1. Construct new atoms and partition them into disjoint classes in two arbitrary ways, by 
their ``state'' and by their ``switch''.\\
2. Link all atoms of the same state in linear chains, being careful not to close a loop.\\
3. Put all the atoms of the same switch in a likebox.\\
4. Partition the likeboxes into classes, and make one linkbox out of each class by linking all the
likeboxes in a linear chain.\\
\\
A collection of types constructed in this way is said to be in {\em linkbox normal form}, and there is
a nice way of representing it as a diagram.

We draw a circle for every atom, and a dashed lines that connect the circles of the same state. Each 
connected component of the dashed-line graph is a {\em state} of the diagram. We draw solid line between
circles of the same switch, and each clump of circles connected by solid lines is a {\em switch}.
Finally, we draw a rectangle around each class of switches linked together, in such a way that
the rectangle surrounds all the solid lines in the switches it contains, but none of the dashed lines.
This rectangle is a linkbox.

For example, if I form atoms {\tt a'-a''-a'''-a'''' b'-b''-b''' c'-c'' d e f g h} linked as indicated,
and make two linkboxes {\tt P=\{d b' a'\}-\{e c' c'' a''\}} and \linebreak
{\tt Q=\{f a''' a''''\}-\{g b''\}-\{h b'''\}} the corresponding diagram is:
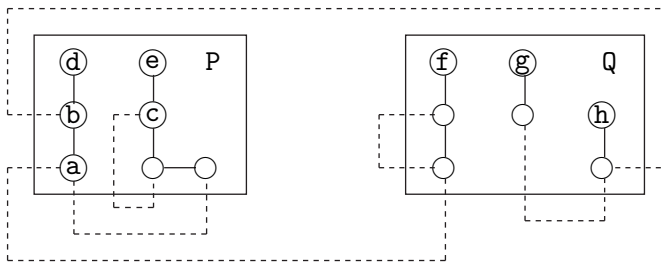
\begin{center}
\begin{picture}(340,100)

\Boxc(75,60)(80,60)\put(100,77){\tt P}
\put(50,80){\circle{10}}\put(47,77){\tt d}\Line(50,75)(50,64)\put(50,60){\circle{10}}\put(47,57){\tt b}\Line(50,55)(50,45)
\put(50,40){\circle{10}}\put(47,38){\tt a}
\put(80,80){\circle{10}}\put(77,78){\tt e}\Line(80,75)(80,65)\put(80,60){\circle{10}}\put(77,58){\tt c}\Line(80,55)(80,44)
\put(80,40){\circle{8}}
\Line(84,40)(96,40)\put(100,40){\circle{8}}

\DashLine(75,60)(65,60){2}\DashLine(65,60)(65,25){2}\DashLine(65,25)(80,25){2}\DashLine(80,25)(80,34){2}
\DashLine(50,35)(50,15){2}\DashLine(50,15)(100,15){2}\DashLine(100,15)(100,34){2}

\Boxc(220,60)(90,60)\put(250,77){\tt Q}
\put(190,80){\circle{10}}\put(187,77){\tt f}\Line(190,75)(190,64)\put(190,60){\circle{8}}\Line(190,56)(190,44)\put(190,40){\circle{8}}
\put(220,80){\circle{10}}\put(217,78){\tt g}\Line(220,75)(220,64)\put(220,60){\circle{8}}
\put(250,60){\circle{10}}\put(247,57){\tt h}\Line(250,55)(250,44)\put(250,40){\circle{8}}

\DashLine(186,60)(165,60){2}\DashLine(165,60)(165,40){2}\DashLine(165,40)(186,40){2}
\DashLine(190,36)(190,5){2}\DashLine(190,5)(25,5){2}\DashLine(25,5)(25,40){2}\DashLine(25,40)(45,40){2}
\DashLine(220,56)(220,20){2}\DashLine(220,20)(250,20){2}\DashLine(250,20)(250,36){2}
\DashLine(45,60)(25,60){2}\DashLine(25,60)(25,100){2}\DashLine(25,100)(275,100){2}\DashLine(275,100)(275,40){2}
\DashLine(275,40)(254,40){2}
\end{picture}
\end{center}

The diagram representation is better because we don't have to give everything a name. In the diagram,
the state-circle labeled {\tt a} corresponds to the atom named {\tt a'}. The circles you reach from
{\tt a} by following dotted lines are {\tt a''},{\tt a'''} and {\tt a''''}. Likewise for the circle
labeled {\tt b}, which includes the {\tt b} family of atoms, and the circle labeled {\tt c}, which
includes the {\tt c}s. The circles which occur unconnected to dotted lines are {\em unbound states},
and they are {\tt d} {\tt e} {\tt f} {\tt g} and {\tt h}.

The picture also suggests a molecular interpretation--- each box is a different kind of molecule, each
switch is a binding domain of this molecule, and each circle a binding state of the domains that
it appears in. Circles connected by dotted lines are connected states, which in the diagram are thought
of as one shared state. For this reason, the dotted lines are called {\em equivalence lines}.

The kinematics of this example is complicated, because it was chosen to illustrate the most general
case.  I will understand the kinematics of simpler examples first, and then come back to it. The
first two examples are familiar:
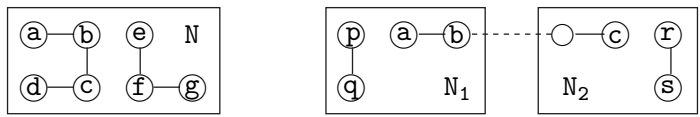
\begin{center}
\begin{picture}(280,80)
\Boxc(40,40)(80,40)\put(67,47){\tt N}
\put(10,50){\circle{10}}\put(7,48){\tt a}\Line(15,50)(25,50)\put(30,50){\circle{10}}\put(27,47){\tt b}
\Line(30,45)(30,35)\put(30,30){\circle{10}}\put(27,28){\tt c}\Line(25,30)(15,30)\put(10,30){\circle{10}}\put(7,27){\tt d}
\put(50,50){\circle{10}}\put(47,48){\tt e}\Line(50,45)(50,35)\put(50,30){\circle{10}}\put(47,27){\tt f}
\Line(55,30)(65,30)\put(70,30){\circle{10}}\put(67,28){\tt g}
\Boxc(150,40)(60,40)\put(165,27){${\tt N_1}$}
\put(130,50){\circle{10}}\put(128,48){\tt p}\Line(130,45)(130,35)\put(130,30){\circle{10}}\put(127,28){\tt q}
\put(150,50){\circle{10}}\put(147,48){\tt a}\Line(155,50)(165,50)\put(170,50){\circle{10}}\put(167,47){\tt b}
\DashLine(175,50)(205,50){2}
\Boxc(230,40)(60,40)\put(210,27){${\tt N_2}$}
\put(210,50){\circle{8}}\Line(214,50)(225,50)\put(230,50){\circle{10}}\put(227,47){\tt c}
\put(250,50){\circle{10}}\put(247,48){\tt r}\Line(250,45)(250,35)\put(250,30){\circle{10}}\put(247,28){\tt s}
\end{picture}
\end{center}
These represent the molecule with twelve internal states and the pair of molecules which bind. The next two
examples are also from the previous section, but this time I will omit the state names.
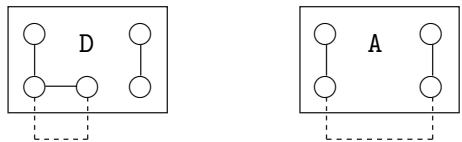
\begin{center}
\begin{picture}(280,80)
\Boxc(40,40)(60,40)\put(37,43){\tt D}
\put(20,50){\circle{8}}\Line(20,46)(20,34)\put(20,30){\circle{8}}\Line(24,30)(36,30)\put(40,30){\circle{8}}
\put(60,50){\circle{8}}\Line(60,46)(60,34)\put(60,30){\circle{8}}
\DashLine(20,26)(20,10){2}\DashLine(20,10)(40,10){2}\DashLine(40,10)(40,26){2}

\Boxc(150,40)(60,40)
\put(130,50){\circle{8}}\Line(130,46)(130,34)\put(130,30){\circle{8}}
\put(170,50){\circle{8}}\Line(170,46)(170,34)\put(170,30){\circle{8}}
\DashLine(130,26)(130,10){2}\DashLine(130,10)(170,10){2}\DashLine(170,10)(170,26){2},
\put(143,43){\tt A}

\end{picture}
\end{center}
{\tt D} is the first-domain dimer, {\tt A} is the linear polymer.

When I don't explicitly name states in the same switch, I will put them in a definite arbitrary
order, so that the index is well defined. The order is from left to right and top to bottom, like
English text, with the exception that more highly linked states come after sparsely linked ones.
States which link four domains come after states which link three, and unbound states come first
of all. The order of the switches in a linkbox is by the English-language order of state-circle
zero, and the order of anonymous linkboxes is by the English language order of their top-left
point.

We describe a configuration of a linkbox is by a text string which has the linkbox name followed
by a list of numbers, e.g. {\tt A<1,1>}. The numbers may then be followed by a hyphen and a plus
separated list of more linkbox names, each of which can be followed by more numbers.  One of the
configurations of {\tt D} is {\tt D<0,0>}, which is {\tt D} unbound. {\tt D<1-D<2,0>,1>} represents
the formed dimer, with the two monomers in different phosphorylation states. {\tt A<0,1-A<1,1-A<1,1-A<1,0>>>>}
is a fourfold bound {\tt A}. One configuration of the {\tt P-Q} diagram is:
\begin{tt}
\begin{tabbing}
++++++\=+++\=+++++\=++++++++\kill
Q<\>   1 - Q<2,0,0> + P<2,1-P<0,2>> +  P<0,3>,\\
\>     1-Q<0,0,1>+P<1,2-P<0,1>>,\\
\>     0\\
\ >
\end{tabbing}
\end{tt}

One nice thing about this representation is that I can omit the internal names without ambiguity, e.g.
{\tt A<0,1-<1,1-<1,1-<1,0>>>}, and omit certain numbers which are demanded by consistency, e.g. 
{\tt A<0,<,<,<,0>>>>}, {\tt D<1-<,0>,1>}. This is fine as long as I can unambiguously reconstruct
the molecules. {\tt D<<,0>,1>}, for example, is {\em not} allowed, because this could be interpreted
as either {\tt D<1-<2,0>,1>} or {\tt D<2-<1,0>,1>}.

To show that the description is useful, the next examples are drawn randomly from models implicitly
described in the biology literature.

The first one describes the binding of {\tt EGFR}, a receptor which binds {\tt EGF}. The model
is that {\tt EGF} binds {\tt EGFR}, then dimerizes and autophosphorylates. A kinematic diagram
for this process is:
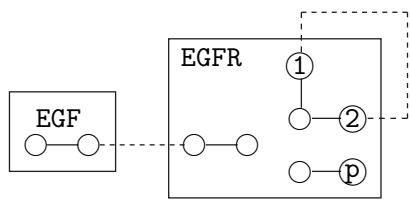
\begin{center}
\begin{picture}(240,80)

\Boxc(40,35)(40,30)\put(30,37){\tt EGF}
\Boxc(120,40)(80,60)\put(85,60){\tt EGFR}

\put(30,30){\circle{8}}\Line(34,30)(46,30)\put(50,30){\circle{8}}\DashLine(54,30)(86,30){2}
\put(90,30){\circle{8}}\Line(94,30)(106,30)\put(110,30){\circle{8}}
\put(130,20){\circle{8}}\Line(134,20)(145,20)\put(150,20){\circle{10}}\put(147,18){\tt p}
\put(130,40){\circle{8}}\Line(130,44)(130,55)\put(130,60){\circle{10}}
\Line(134,40)(145,40)\put(150,40){\circle{10}}\DashLine(155,40)(170,40){2}\DashLine(170,40)(170,80){2}
\DashLine(170,80)(130,80){2}\DashLine(130,80)(130,65){2}
\put(127,57){\tt 1}\put(147,37){\tt 2}

\end{picture}
\end{center}
Whether this story is physically true in the details is not so important. What is important
is that it predicts the behavior of {\tt EGF} in a wide variety of signaling experiments,
and therefore captures the function with a minimum number of bits.

The active states of {\tt EGFR} are {\tt EGFR<1,1,p>} and {\tt EGFR<2,1,p>}, left and right
phosphorylated monomers components of a dimer. The states {\tt 1} and {\tt 2} of the
dimerization domain are numbered in English language order, but to help the new reader, I put
the numbers in. {\tt EGFR<1-<2,1,1>,1,1>} or {\tt EGFR<1-<2,1,p>,1,p>} is a doubly active
complex.

The second diagram describes a model of {\tt c-Abl/v-Abl} activation of {\tt Ras}, implicitly
described in \cite{Linda}. {\tt c-Abl} is a cell signaling protein with known activity on both
{\tt p73} and {\tt Shc}, while {\tt v-Abl} is a variant expressed during viral infection.
This diagram includes some dynamical information, but it is easy to reconstruct the kinematics---
replace each vertex by a circle, replace every squiggly line by a solid line, and ignore the lines
that ends in `$>$' and `x'. The kinematics of the switch in {\tt Shc} guarantees that the binding of the
two variants of {\tt abl} is competitive. There is one construction which is not in normal form---
{\tt GRB/SOS} is a diagrammatic atom. It can be interpreted as a linkbox with one switch that
binds {\tt Shc}.
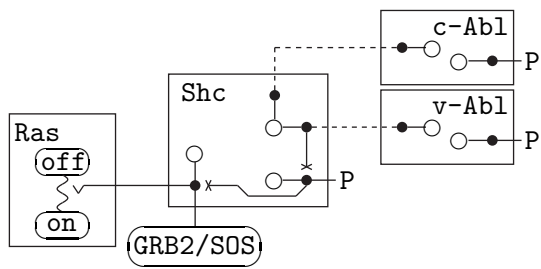
\begin{center}
\begin{picture}(240,110)

\Boxc(90,55)(60,50)\put(65,70){\tt Shc}
\put(100,60){\circle{6}}\Line(103,60)(110,60)\Vertex(112,60){2}\DashLine(114,60)(146,60){2}\Vertex(148,60){2}
\Line(150,60)(157,60)\put(160,60){\circle{6}}

\Boxc(165,90)(50,28)\put(160,95){\tt c-Abl}
\Boxc(165,60)(50,28)\put(160,65){\tt v-Abl}
\put(170,55){\circle{6}}\Line(173,55)(193,55)\Vertex(182,55){2}\put(195,52){\tt P}
\put(170,85){\circle{6}}\Line(173,85)(193,85)\Vertex(182,85){2}\put(195,82){\tt P}

\Line(112,58)(112,45)\Line(110,46)(114,44)\Line(110,44)(114,46)
\Line(112,38)(108,34)\Line(108,34)(90,34)\Line(90,34)(86,38)\Line(86,38)(75,38)\Line(76,40)(74,36)\Line(76,36)(74,40)

\Line(100,63)(100,70)\Vertex(100,72){2}\DashLine(100,74)(100,90){2}\DashLine(100,90)(146,90){2}\Vertex(148,90){2}
\Line(150,90)(157,90)\put(160,90){\circle{6}}

\put(100,40){\circle{6}}\Line(103,40)(123,40)\Vertex(112,40){2}\put(125,37){\tt P}

\put(70,50){\circle{6}}\Line(70,47)(70,22)\Vertex(70,38){2}\put(70,15){\oval(50,15)}
\put(47,12){\tt GRB2/SOS}
\Line(70,38)(28,38)\Line(28,38)(26,35)\Line(26,35)(24,38)

\Boxc(20,40)(40,50)\put(2,55){\tt Ras}
\put(20,47){\oval(20,10)}\put(12,44){\tt off}
\Photon(20,42)(20,28){2}{2}
\put(20,23){\oval(20,10)}\put(14,21){\tt on}

\end{picture}
\end{center}
A diagram can be drawn in many ways--- the function is a function of the model. The {\tt Abl} diagram
in particular has different form, in which the {\tt c-Abl} and {\tt v-Abl} linkboxes are combined. This
is the form I will use to describe the in-vitro experiments described in \cite{Israelis}, which studied
the differential binding of {\tt c-Abl} and {\tt v-Abl} to {\tt RFXI} on the {\tt p73} binding domain.
This diagram is derived from an in-vitro experiment, so it contains more detailed information.
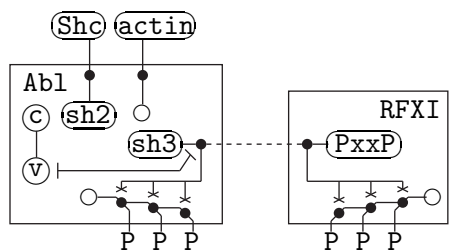
\begin{center}
\begin{picture}(240,100)
\Boxc(70,40)(80,60)\put(35,60){\tt Abl}
\put(40,50){\circle{10}}\put(37,48){\tt c}\Line(40,45)(40,35)\put(40,30){\circle{10}}\put(37,28){\tt v}
\put(60,51){\oval(20,10)}\put(51,48){\tt sh2}\Line(60,56)(60,80)\Vertex(60,66){2}\put(56,85){\oval(20,10)}\put(48,82){\tt Shc}
\put(80,52){\circle{6}}\Line(80,55)(80,80)\Vertex(80,66){2}\put(85,85){\oval(30,10)}\put(71,82){\tt actin}
\put(85,40){\oval(20,10)}\put(76,37){\tt sh3}\Line(95,40)(100,40)\Vertex(102,40){2}\DashLine(104,40)(140,40){2}
\Vertex(142,40){2}\Line(144,40)(150,40)\put(164,40){\oval(28,10)}\put(153,37){\tt PxxP}
\Boxc(165,35)(60,50)\put(170,50){\tt RFXI}
\Line(48,30)(94,30)\Line(94,30)(98,36)\Line(100,34)(96,38)\Line(48,27)(48,33)

\put(60,20){\circle{6}}\Line(63,20)(70,20)\Line(70,20)(75,15)\Vertex(72,18){2}\Line(75,15)(75,7)\put(72,0){\tt P}
\Line(72,18)(82,18)\Line(82,18)(87,13)\Line(87,13)(87,7)\Vertex(84,16){2}\put(84,0){\tt P}
\Line(84,16)(94,16)\Line(94,16)(99,11)\Line(99,11)(99,7)\Vertex(96,14){2}\put(96,0){\tt P}

\Line(102,40)(102,26)\Line(102,26)(72,26)\Line(72,26)(72,23)\Line(70,22)(74,24)\Line(70,24)(74,22)
\Line(84,26)(84,21)\Line(82,22)(86,20)\Line(82,20)(86,22)
\Line(96,26)(96,19)\Line(94,20)(98,18)\Line(94,18)(98,20)

\put(190,20){\circle{6}}\Line(187,20)(180,20)\Line(180,20)(175,15)\Vertex(178,18){2}\Line(175,15)(175,7)
\put(173,0){\tt P}
\Line(178,18)(168,18)\Line(168,18)(163,13)\Line(163,13)(163,7)\Vertex(166,16){2}\put(161,0){\tt P}
\Line(166,16)(156,16)\Line(156,16)(151,11)\Line(151,11)(151,7)\Vertex(154,14){2}\put(149,0){\tt P}

\Line(142,40)(142,26)\Line(142,26)(178,26)\Line(178,26)(178,23)\Line(176,24)(180,22)\Line(176,22)(180,24)
\Line(166,26)(166,21)\Line(164,22)(168,20)\Line(164,20)(168,22)
\Line(154,26)(154,19)\Line(152,20)(156,18)\Line(152,18)(156,20)

\end{picture}
\end{center}
This diagram also contains a new kinematic construction: a {\em kinematic inhibition}. This is the line
ending in vertical bars. A kinematic inhibition forbids two states from occurring together in the same
molecule. This construction can almost be translated into normal form, and I will discuss how to do this
in the next to last section. The chains of phosphorylation also need comment, since they do not consider the
independent phosphorylations, only the net phosphorylation state. This is an approximation, accurate
if the function of a molecule only depends on the net charge. The information capacity of a molecule
whose function is only sensitive to net charge is much smaller than one whose function depends on the
detailed pattern of phosphorylation.

In principle, these diagrams should not be combined, since the models come from different sources. I
will do so anyway, and add a little bit more function on {\tt p73} \cite{Agami}. The larger diagram
may not be a correct model of anything anymore, but it is useful for qualitative prediction of function,
as a starting point for a quantitative model, and for analyzing the information processing capacity of
protein networks.
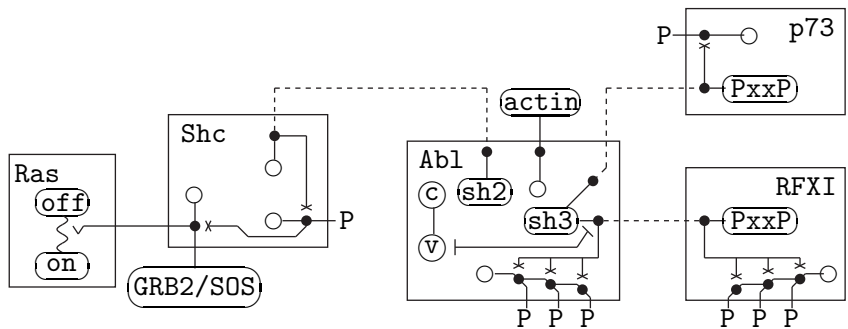
\begin{center}
\begin{picture}(340,120)

\Boxc(90,55)(60,50)\put(65,70){\tt Shc}
\put(100,60){\circle{6}}

\Line(100,72)(112,72)\Line(112,72)(112,45)\Line(110,46)(114,44)\Line(110,44)(114,46)
\Line(112,38)(108,34)\Line(108,34)(90,34)\Line(90,34)(86,38)\Line(86,38)(75,38)\Line(76,40)(74,36)\Line(76,36)(74,40)

\Line(100,63)(100,70)\Vertex(100,72){2}\DashLine(100,74)(100,90){2}\DashLine(100,90)(180,90){2}

\put(100,40){\circle{6}}\Line(103,40)(123,40)\Vertex(112,40){2}\put(125,37){\tt P}

\put(70,50){\circle{6}}\Line(70,47)(70,22)\Vertex(70,38){2}\put(70,15){\oval(50,15)}
\put(47,12){\tt GRB2/SOS}
\Line(70,38)(28,38)\Line(28,38)(26,35)\Line(26,35)(24,38)

\Boxc(20,40)(40,50)\put(2,55){\tt Ras}
\put(20,47){\oval(20,10)}\put(12,44){\tt off}
\Photon(20,42)(20,28){2}{2}
\put(20,23){\oval(20,10)}\put(14,21){\tt on}

\Boxc(190,40)(80,60)\put(155,60){\tt Abl}
\put(160,50){\circle{10}}\put(157,48){\tt c}\Line(160,45)(160,35)\put(160,30){\circle{10}}\put(157,28){\tt v}
\put(180,51){\oval(20,10)}\put(171,48){\tt sh2}\Line(180,56)(180,66)\Vertex(180,66){2}\DashLine(180,66)(180,90){2}
\put(200,52){\circle{6}}\Line(200,55)(200,80)\Vertex(200,66){2}\put(201,85){\oval(30,10)}\put(187,82){\tt actin}
\put(205,40){\oval(20,10)}\put(196,37){\tt sh3}\Line(215,40)(220,40)\Vertex(222,40){2}\DashLine(224,40)(260,40){2}
\Vertex(262,40){2}\Line(264,40)(270,40)\put(284,40){\oval(28,10)}\put(273,37){\tt PxxP}
\Boxc(285,35)(60,50)\put(290,50){\tt RFXI}
\Line(168,30)(214,30)\Line(214,30)(218,36)\Line(220,34)(216,38)\Line(168,27)(168,33)

\Line(210,45)(220,55)\Vertex(220,55){2}\DashLine(220,55)(225,60){2}\DashLine(225,60)(225,90){2}
\DashLine(225,90)(262,90){2}\Vertex(262,90){2}\Line(264,90)(270,90)\put(284,90){\oval(28,10)}\put(273,87){\tt PxxP}
\Boxc(285,100)(60,40)\put(295,110){\tt p73}
\put(280,110){\circle{6}}\Line(277,110)(250,110)\Vertex(262,110){2}\put(245,107){\tt P}
\Line(262,90)(262,106)\Line(260,107)(264,105)\Line(260,105)(264,107)

\put(180,20){\circle{6}}\Line(183,20)(190,20)\Line(190,20)(195,15)\Vertex(192,18){2}\Line(195,15)(195,7)\put(192,0){\tt P}
\Line(192,18)(202,18)\Line(202,18)(207,13)\Line(207,13)(207,7)\Vertex(204,16){2}\put(204,0){\tt P}
\Line(204,16)(214,16)\Line(214,16)(219,11)\Line(219,11)(219,7)\Vertex(216,14){2}\put(216,0){\tt P}

\Line(222,40)(222,26)\Line(222,26)(192,26)\Line(192,26)(192,23)\Line(190,22)(194,24)\Line(190,24)(194,22)
\Line(204,26)(204,21)\Line(202,22)(206,20)\Line(202,20)(206,22)
\Line(216,26)(216,19)\Line(214,20)(218,18)\Line(214,18)(218,20)

\put(310,20){\circle{6}}\Line(307,20)(300,20)\Line(300,20)(295,15)\Vertex(298,18){2}\Line(295,15)(295,7)
\put(293,0){\tt P}
\Line(298,18)(288,18)\Line(288,18)(283,13)\Line(283,13)(283,7)\Vertex(286,16){2}\put(281,0){\tt P}
\Line(286,16)(276,16)\Line(276,16)(271,11)\Line(271,11)(271,7)\Vertex(274,14){2}\put(269,0){\tt P}

\Line(262,40)(262,26)\Line(262,26)(298,26)\Line(298,26)(298,23)\Line(296,24)(300,22)\Line(296,22)(300,24)
\Line(286,26)(286,21)\Line(284,22)(288,20)\Line(284,20)(288,22)
\Line(274,26)(274,19)\Line(272,20)(276,18)\Line(272,18)(276,20)

\end{picture}
\end{center}
This process can be iterated to produce maps of the function of thousands of proteins,
in various levels of detail.

Now I will come back to the arbitrary example which began the section. This diagram describes two molecules
{\tt P} and {\tt Q}. {\tt P} polymerizes through a binding of its first and second domain, and every
adjacent pair in the polymer is necessarily also attached to a pair of {\tt Q} molecules, dimerized
in their first domain. Each {\tt Q} molecule has a second and third domain, and it can polymerize by
binding them together. Every pair of polymerized {\tt Q} molecules must be joined to a {\tt P},
unpolymerized in its first domain, but possibly polymerized or dimerized in its second domain.
Dimerized? Yes, because any {\tt P} unpolymerized in its second domain can dimerize by a binding
of this domain to itself. Although it is hard to express the patterns of binding in English, the
diagram is small and clear.

\section{Syntax and Semantics of Diagrams}

The {\em syntax} of a kinematic diagrams is a rule which tells us which diagrams are allowed. A
syntactical kinematic diagram is any collection of nonoverlapping circles and boxes, lines and
dotted lines, such that each line connects two circles entirely inside a box, and each dotted
line connects two circles and is not entirely inside a box. 

The semantics of a diagram is the collection of all molecular complexes it describes. I
must define the semantics carefully, because if I make {\tt D<1-D<2,0>,1>} the same as
{\tt D<2-D<1,0>,1>}, there is no efficient algorithm for listing all the states. It
is not a good idea to define the semantics using the resolution strings, because some
strings, like {\tt D<1-D<2-D<1,1>,0>,0>}, don't make any sense.

I will define the complexes of a diagram as a collection of directed graphs with ordered edges,
which are easiest to understand as mutually pointing objects\\
\\
A {\em molecule} is a pointer to a linkbox {\em type}, followed by a list of stats from
each switch in turn.\\
\\
A {\em stat} is a pointer to a state, followed by a list of pointers to {\em parent} molecules.\\
\\
These must obey three consistency conditions:\\
\\
1. The $k$-th stat in a molecule of type {\tt L} must be of a type that occurs in the $k$-th switch of {\tt L}.\\
2. The $k$-th parent of a stat of type {\tt S} must be of the type of the owner of the $k$-th switch containing
{\tt S}.\\
3. The pointers must come in back-forth pairs: If a configuration for state $S$ points to a
molecule in position $k$, the molecule must point to the configuration at the position of the
corresponding switch. Equivalently, if a molecule points to a configuration in switch $k$,
the configuration must point to the molecule at the position of that switch. This is only
slightly ambiguous when the state occurs more than once in the same switch, in which case
the order of the back-pointers is the order of occurrence of the state-circles of the state.\\
\\
There is one additional condition--- non-circularity.\\
\\
NC: No sequence of molecule-stat pointer hops can end on the same molecule without backtracking.\\
\\
Two complexes are {\em equivalent} when there is a one-to-one correspondence between the molecules
and stats of the two which preserves the pointer structure. More precisely, a map {\em respects
pointing} if whenever $q$ points to $r$ in positions $k_1 ... k_l$, the image of $q$ points
to the image of $r$ in the same places. An equivalence is a bijection between complexes that
takes molecules to molecules and stats to stats and respect pointing. This is an isomorphism
of directed graphs which preserves order of edges.\\
\\
The equivalence classes of finite complexes obeying NC define the noncircular kinematics of
a diagram. This completes the tree semantics.\\
\\
The {\em index} of a stat in a molecule is defined as the number of the stat's type in a switch.
If the state occurs more than once, this is ambiguous, and the ambiguity is resolved by the
back-pointer order in the stat. If {\tt l} contains {\tt s} in position $k$, the index of
$s$ is determined by where the back-pointer to $l$ occurs.

I will make no distinction between complexes and their equivalence classes. This means that
I should check that future definitions respect equivalence, but I won't since it is usually
obvious. Complexes which don't obey NC are called {\em circular}, and the {\em finite complexes}
of a diagram include both the circular and non-circular ones.\\
\\
I will extend complexes a little bit, so as to describe molecules in infinite chains. This is
useful because some diagrams only have infinite configurations.\\
\\
A {\em partial complex} is a complex (possibly circular) with {\em null} pointers in the place
of some of the molecule or stat pointers. A null pointer cannot be dereferenced, and passes
all consistency checks. This defines which partial complexes are allowed.\\
\\
Observe that every pointer is checked for consistency against exactly one other pointer, this is
the content of condition 4, so I can insist that only null pointers have null partners without
loss of generality.\\
\\
A partial complex {\tt C'} {\em extends} partial complex {\tt C} if {\tt C'} contains {\tt C} as a
subcomplex, except with some null pointers now non-null.\\
\\
Proposition: Every consistent partial complex can be consistently extended.\\
\\
Proof: Replace the null pointer and its null partner with a pointer to a new object of the right
type.\\
\\
Two partial complexes with a stat of the same type can be {\em attached} if the non-null parent
pointers of the two stats are disjoint. Attaching deletes one of the two configurations, and
merges the two pointer lists. This can be considered an extension of either complex.\\
\\
Similarly, if two partial complexes both have a molecule of the same type with disjoint places
where the configuration pointer is non-null, you can merge the two molecules into one, and this
is how you attach molecules. Attachment on a given set of pointers is associative and commutative---
it doesn't depend on the order of attachment--- so long as you don't try to attach two different things
at the same spot.\\ 
\\
Two partial complexes are {\em equivalent} if there is a one-to-one correspondence between their
molecules and configurations which respects pointing. This map must take null pointers to null pointers.\\
\\
I will think of a partial complex as representing the collection of its extensions. A partial
complex {\tt A} {\em contains} {\tt B} if {\tt B} is an extension of {\tt A}. Two partial
complexes {\em overlap} if they have an equivalent extension.\\
\\
Using partial complexes, I can define infinite complexes as a limit. An infinite complex is a sequence
of partial complex, each extending the previous one. An extending sequences of partial complexes {\tt C}
{\em interleaves} extending sequences {\tt A} and {\tt B} if an infinite subsequence of {\tt A} and an
infinite subsequence of {\tt B} both occur as infinite subsequences of {\tt C}. Two infinite extending
sequences define the same infinite complex when they can be interleaved.\\
\\
The cardinality of the set of infinite complexes is very easily that of the continuum. This means
that their combinatorics depends on the model of set theory you like to use. It is therefore best
to frame all questions in a way that is computational and absolute.\\\ 
\\
I will now define the language for resolutions, first in informal mathematics, then as a BNF.\\
\\
A {\em linkbox resolution} is a linkbox name followed by a list of comma-separated switch
resolutions, one for each switch in the linkbox.\\
\\
A {\em switch resolution} is an integer index $i$ between 0 and the size of the switch,
followed by an optional hyphen and plus separated list of linkbox resolutions.\\
\\
These definitions form a recursive pair, which given a diagram define a language of text strings.
The consistency conditions are:\\
\\
R1: The number and type of linkbox resolutions in a switch resolutions are one less than
the number and type of linkboxes that the i-th state of the switch is linked to by dotted
lines. This is one less than the number of places the state circle appears.\\
R2: The number and type of switch resolutions in a linkbox resolutions are the number of
switches in the linkbox.\\
\\
The BNF of resolutions restates the informal definition, but it defines how to build a tree from
a resolution using any standard parse tree construction algorithm. Here it is: 
\begin{tt}
\begin{tabbing}
+++++++++++++++++++++++++++\=\kill
\verb@linkbox_resolution_list:@	\> \verb@linkbox_resolution_list@ | \\
				\> \verb@linkbox_resolution_list `+' linkbox_resolution@\\
\verb@linkbox_resolution:@ 	\> \verb@LINKBOX [`<' state_resolution_list `>']@\\
\verb@state_resolution_list:@	\> \verb@state_resolution_list@ | \\
				\> \verb@state_resolution_list `,' state_resolution@\\
\verb@state_resolution:@	\> \verb@[ INTEGER [`-' linkbox_resolution_list]]@\\
\end{tabbing}
\end{tt}
Square brackets surround optional constructions. The tokens are linkbox names and integers.
This is everything, except for a little nesting context sensitivity.  Inside angle-brackets,
we remember the linkbox we are resolving, and as we pass each comma, we keep track of the
switch.\\
\\
A resolution {\em builds} a partial complex $C$ following the BNF parse tree construction.
When this algorithm fails the resolution is {\em inconsistent}, and this condition defines
the resolution syntax. All algorithms act by filling in pointers of partial complexes, and
the end result can be partial. The partial complex corresponding to the top node of the parse
tree is the meaning of a resolution string, and since partial complexes are identified with
the collection of all their extensions, resolutions naturally define {\em classes} of complexes.\\
\\
The object constructed at each node is described below:
\\
{\tt linkbox\_resolution\_list:} makes a list from the molecules that its second rule eats.\\
{\tt linkbox\_resolution:} creates a new molecule of type the linkbox, and attaches it to the
stat list from the {\tt state\_resolution\_list} node below checking R1. If there is no
list, the molecule is null.\\
{\tt state\_resolution\_list:} makes a list of stats from its second rule.\\
{\tt state\_resolution:} creates a stat of type the $i$-th state of the current switch, where
$i$ is the integer argument. If there is no integer, it builds a null state. If there
is a {\tt linkbox\_resolution\_list}, it attaches the state to the molecules in the list
checking R2.\\
\\
Observe that the context is only used to figure out what the integers mean. If we were to name every
state-circle, and identify the state-circles by names instead of numbers, the grammar would be context
free.\\
\\
Proposition: The resolution syntax constructs every noncircular partial complex.\\
\\
Proof: I will work with the context-free form. The parse tree of the grammar can be any tree with
any amount of branching, and the branching occurs right below the {\tt linkbox\_resolution} and
{\tt state\_resolution} nodes. The build algorithm builds a link-stat tree of the same shape as
the parse tree, since it attaches every level to the one below. At any node, I can put a null
molecule or a null stat, since the rules for {\tt linkbox\_resolution} and {\tt state\_resolution}
include these as options. The attaching operation adds pointers in back-and-forth pairs--- pointers
are not allowed to overwrite each other during an attaching. So we automatically satisfy condition 4.
This excludes strings like {\tt D<2-D<1-D<2,0>,0>,0>} which could conceivably make sense.\\
\\
To build circular complexes, we must allow the resolutions to contain local variables which can be
used in place of molecules or stats. This means that any identifier may be used in a resolution.  The
identifiers are introduced either by a name alone or by a name followed by an equal sign.\\
\\
The variables are set equal to the resolution constructed at the deepest node where they occur.
They are extended wherever they occur again in the parse tree, using the same algorithms that
attach other partial complexes. If a pointer is ever overwritten with a different value, the
resolution is inconsistent.  The expression {\tt r=A<1,1-r>}, for example, describes a molecule 
of the linear polymer which tried to polymerize, but ended up binding to itself.
{\tt r=D<2-D<1-r=D<2,s=0>,1>,s=0>} is now a consistent, but comically redundant, resolution of the
dimer.\\
\\
Since adding matching variables in two places in a tree closes a loop, the semantics of resolutions
with variables contains all partial complexes. Any graph can be constructed from a spanning tree with
a little identification.

\section{Geometric Formulation}

Although I have chosen to define the semantics of the diagrams computationally, there is an
equivalent geometric definition which is sometimes more intuitive. Here, the molecular complexes
are a certain type of branching path in a diagram. This definition is elegant, but it makes
the equivalence relation on complexes awkward.\\
\\
A {\em state-path} is a finite nonempty sequence of pairs of state circles
${\tt (A_0,B_0),(A_1,B_1),...}$ such that ${\tt B_i}$ is joined to ${\tt A_i}$ by a dotted line
path, and ${\tt A_{i+1}}$ is a circle of the same linkbox as ${\tt B_i}$ but not joined to ${\tt A_i}$
by a solid line path, i.e. ${\tt A_{i+1}}$ and ${B_{i}}$ are in different switches.\\
\\
A {\em circular state-path} is formed from a state-path of length $N$ where ${\tt B_{N-1}}$ is
in the same linkbox as ${\tt A_0}$, but in a different switch. It is a state-path that stays a
state-path after cyclic permutation. We can {\em cut} a circular path in $N$ different ways to
make a linear path.\\
\\
Every state-path determines a noncircular partial complex in a natural way: Build an empty
molecule of the linkbox containing ${\tt A_0}$, attach it to a configuration in the switch-position
of state-circle ${\tt A_0}$, attach this to a new molecule of the linkbox containing ${\tt B_0}$, etc.
This maps a state-path to a unique partial complex.\\
\\
We can do the same thing for circular state-paths, but then we link the last molecule to the first.
This maps a circular state-path to a circular partial complex unique up to cyclic permutation.\\
\\
A state-path {\tt L'} {\em extends} {\tt L} if {\tt L} is an initial subsequence of {\tt L'}. A
circular state-path {\em extends} a non-circular state path if a cyclic permutation either extends
the non-circular path or coincides with it.\\
\\
Proposition: Extending state-paths extends the partial complex.\\
\\
State-paths correspond to connected lines of molecules, but we want trees, so the next thing to do
is allow the state-paths to branch. There are many ways to extend a state paths from ${\tt B_i}$,
as many ways as there are likeboxes in ${\tt B_i}$'s linkbox. Likewise, we can extend a state-path
many ways from {\tt A}, as many ways as there are circles joined to ${\tt A_i}$ by dotted lines.
A branching path goes on more than one path simultaneously.\\
\\
A {\em branching} is a pair of inequivalent extensions of the same path, neither of which extends
the other. If the two paths have different starting points, the two starting points must be in
different switches of the same linkbox.\\
\\
Proposition: Two branching partial complexes can be attached at a unique point.\\
Proposition: We can consider any linear path as a branching based at any point.\\
Proposition: A cut circular state-path is a branching at any base point.\\
Proposition: If we back-extend a branching from the base-point, it stays a branching.\\
\\
Since a branching chooses different directions from a point, branching linear paths define state-path
trees in the diagram. The two extensions extend different pointers of the final molecule of the state
path they branch off of, so we can build a partial complex from any branching.\\
\\
A {\em state-tree} is a collection of noncircular state-paths where each pair makes a branch.\\
\\
Proposition: Every state-tree determines a unique partial complex whose extensions are the intersections
of the partial complex classes defined by each state path.\\
\\
Proof: Attachment does not depend on order, so I will define the partial complex as the attachment
of all the partial complexes corresponding to each path. If we cannot attach a partial complex {\tt P},
this implies that it attaches at a non-null point to a new type, which implies that some other path is
non-null of a different type at this point, and so does not form a branching with {\tt P}.\\
\\
So a non-circular complex can be alternatively defined as a maximal non-circular state-tree, where
a {\em maximal} state-tree is one which is not part of a larger state-tree. The equivalence relation
is inherited from that of partial complexes.\\
\\
A {\em circular complex} is an state-tree with some state-paths back-extensions of cut state-circles.\\
\\
Proposition: Every circular complex defines a unique infinite tree which is it's simply
connected cover.\\
\\
Proof: Unwrap each circular path in the complex to an infinite line.\\
\\
A diagram is {\em recursive} when it contains an infinite complex.\\
\\
Theorem: A diagram is recursive if and only if it contains a closed state-path.\\
\\
Proof: Since a state-path defines a partial complex, we have one direction immediately. Given
a closed state path, traverse it infinitely often to make an extending chain of partial complexes.
The other direction follows from the finite size of a diagram. If there are arbitrarily large
state-trees, there must be arbitrarily long state-paths, since the branching rate is bounded.
Each state-path can only visit finitely many different state-circles without repeating.
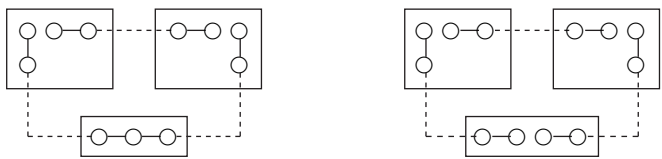
\begin{center}
\begin{picture}(240,80)
\Boxc(22,53)(40,30)
\Boxc(78,53)(40,30)
\put(10,60){\circle{6}}
\Line(10,57)(10,50)\put(10,47){\circle{6}}\DashLine(10,44)(10,20){2}
\put(20,60){\circle{6}}\Line(23,60)(30,60)\put(33,60){\circle{6}}\DashLine(36,60)(64,60){2}
\put(67,60){\circle{6}}\Line(70,60)(77,60)\put(80,60){\circle{6}}
\put(90,60){\circle{6}}\Line(90,57)(90,50)\put(90,47){\circle{6}}\DashLine(90,44)(90,20){2}
\DashLine(90,20)(66,20){2}
\put(50,20){\circle{6}}\Line(53,20)(60,20)\put(63,20){\circle{6}}
\Line(47,20)(40,20)\put(37,20){\circle{6}}\DashLine(34,20)(10,20){2}
\Boxc(50,20)(40,15)

\Boxc(172,53)(40,30)
\Boxc(228,53)(40,30)
\put(160,60){\circle{6}}
\Line(160,57)(160,50)\put(160,47){\circle{6}}\DashLine(160,44)(160,20){2}
\put(170,60){\circle{6}}\Line(173,60)(180,60)\put(183,60){\circle{6}}\DashLine(186,60)(214,60){2}
\put(217,60){\circle{6}}\Line(220,60)(227,60)\put(230,60){\circle{6}}
\put(240,60){\circle{6}}\Line(240,57)(240,50)\put(240,47){\circle{6}}\DashLine(240,44)(240,20){2}
\DashLine(240,20)(221,20){2}
\put(205,20){\circle{6}}\Line(208,20)(215,20)\put(218,20){\circle{6}}
\put(195,20){\circle{6}}\Line(192,20)(185,20)\put(182,20){\circle{6}}\DashLine(179,20)(160,20){2}
\Boxc(200,20)(50,15)

\end{picture}\\
The diagram on the right is recursive, the diagram on the left is not.\\
\end{center}
So a linkbox which contains the same state twice in different switches forms an infinite polymer,
as does a linkbox which can dimerize in two different switches.

\section{Axioms Revisited}

So far, the only thing that I have defined precisely is the normal form, because in this case I
have a diagram and a nice way of representing every complex. What about everything else? What
if I make likeboxes of linkboxes and link those? Or linkboxes of already-existing linkboxes?
It seems that there is a lot more work to do.

It's not true--- I'm done. Because everything that can be built up by the axioms can be built in
normal form, that is, by drawing linkboxes:\\
\\
1. To construct a new atom, draw an empty box.\\
\\
2. To construct a binary likebox of two linkboxes {\tt A} and {\tt B}, join two circles by a line
and put them in a new linkbox. Then join one of the circles by a dotted line to a new circle in
{\tt A}, and the other to a new circle in {\tt B}.\\
\\
3. To construct a binary linkage of two linkboxes {\tt A} and {\tt B}, put a new circle in each box,
and join the new circles by dashed lines to two circles in a box, this time without a solid line
connecting them.\\
\\
In picture form:\\
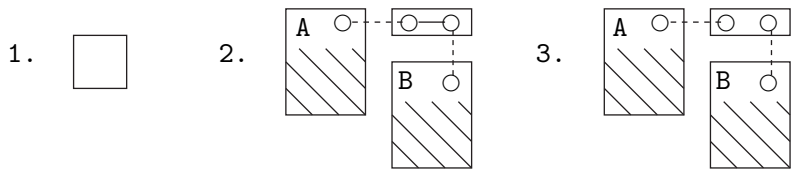
\begin{center}
\begin{picture}(340,60)
\put(0,40){\tt 1.} \Boxc(35,40)(20,20)

\put(80,40){\tt 2.} \Boxc(120,40)(30,40)\put(109,50){\tt A}
\Line(105,30)(115,20)
\Line(105,40)(125,20)
\Line(110,45)(135,20)
\Line(120,45)(135,30)
\Line(130,45)(135,40)
\Boxc(160,20)(30,40)\put(148,30){\tt B}
\Line(145,10)(155,0)
\Line(145,20)(165,0)
\Line(150,25)(175,0)
\Line(160,25)(175,10)
\Line(170,25)(175,20)
\Boxc(160,55)(30,10)
\put(152,55){\circle{6}}\Line(155,55)(165,55)\put(168,55){\circle{6}}
\DashLine(149,55)(130,55){2}\put(127,55){\circle{6}}
\DashLine(168,52)(168,35){2}\put(168,32){\circle{6}}

\put(200,40){\tt 3.} \Boxc(240,40)(30,40)\put(229,50){\tt A}
\Line(225,30)(235,20)
\Line(225,40)(245,20)
\Line(230,45)(255,20)
\Line(240,45)(255,30)
\Line(250,45)(255,40)
\Boxc(280,20)(30,40)\put(268,30){\tt B}
\Line(265,10)(275,0)
\Line(265,20)(285,0)
\Line(270,25)(295,0)
\Line(280,25)(295,10)
\Line(290,25)(295,20)
\Boxc(280,55)(30,10)
\put(272,55){\circle{6}}\put(288,55){\circle{6}}
\DashLine(269,55)(250,55){2}\put(247,55){\circle{6}}
\DashLine(288,52)(288,35){2}\put(288,32){\circle{6}}
\end{picture}\\
\end{center}
It is easy to verify that the kinematics of these diagrams reproduce the intuitive kinematics
that are demanded by the axioms. But every new type constructed by the procedures above gets
its own linkbox, so we can iterate the operations indefinitely, and everything stays in normal
form.

We have pulled ourselves up by the bootstraps. Every collection of atoms, likeboxes, and linkages
is unambiguously defined. It is now easy to check that:\\
\\
Proposition: If we link an atom to itself, all the complexes are infinite (partial).\\ 
Proposition: If we make likeboxes of likeboxes, every link-path passing through the representative
of the top likebox passes through exactly one of the children.\\
Proposition: In any axiomatic normal form, every state is linked to exactly one other one.\\
\\
I can now extend the diagram syntax to include arbitrary likeboxes and equivalence lines. A likebox
is drawn as a round-corner object surrounding the items that it contains.  If the item is a likebox
or a linkbox, the likebox contains the box but not any of the things inside. An empty likebox or linkbox
is an atom, and a linkbox can only contain atomic states. A linkbox cannot contain another linkbox, if
you want that, link the inner linkbox to a state with an equivalence line.

In principle any diagram can be drawn using only likeboxes and equivalence lines, since equivalence lines
are just linkages. Such a diagram is in {\em likebox form}, and the syntax of this form is trivial--- any
collection of arbitrarily nesting likeboxes connected in any pattern. I won't draw biological diagrams this
way, because then they would be incomprehensible. But likebox form makes the logical structure of the
formalism clearer.

If you want to convert linkboxes to likeboxes, do this:
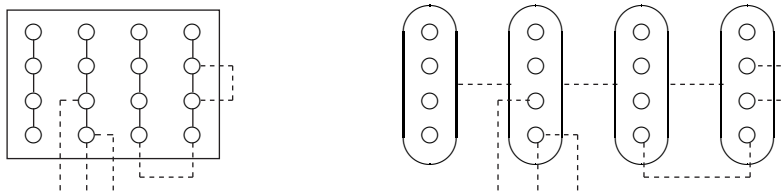
\begin{center}
\begin{picture}(300,80)
\Boxc(40,40)(80,56)
\put(10,60){\circle{6}}\Line(10,57)(10,50)\put(10,47){\circle{6}}\Line(10,44)(10,37)\put(10,34){\circle{6}}
\Line(10,31)(10,24)\put(10,21){\circle{6}}
\put(30,60){\circle{6}}\Line(30,57)(30,50)\put(30,47){\circle{6}}\Line(30,44)(30,37)\put(30,34){\circle{6}}
\Line(30,31)(30,24)\put(30,21){\circle{6}}
\put(50,60){\circle{6}}\Line(50,57)(50,50)\put(50,47){\circle{6}}\Line(50,44)(50,37)\put(50,34){\circle{6}}
\Line(50,31)(50,24)\put(50,21){\circle{6}}
\put(70,60){\circle{6}}\Line(70,57)(70,50)\put(70,47){\circle{6}}\Line(70,44)(70,37)\put(70,34){\circle{6}}
\Line(70,31)(70,24)\put(70,21){\circle{6}}
\DashLine(27,34)(20,34){2}\DashLine(20,34)(20,0){2}
\DashLine(30,18)(30,0){2}
\DashLine(33,21)(40,21){2}\DashLine(40,20)(40,0){2}
\DashLine(73,47)(85,47){2}\DashLine(85,47)(85,34){2}\DashLine(85,34)(73,34){2}
\DashLine(70,18)(70,5){2}\DashLine(70,5)(50,5){2}\DashLine(50,5)(50,18){2}

\put(160,40){\oval(20,60)}\DashLine(170,40)(190,40){2}
\put(160,60){\circle{6}}\put(160,47){\circle{6}}\put(160,34){\circle{6}}
\put(160,21){\circle{6}}
\put(200,40){\oval(20,60)}\DashLine(210,40)(230,40){2}
\put(200,60){\circle{6}}\put(200,47){\circle{6}}\put(200,34){\circle{6}}
\put(200,21){\circle{6}}
\put(240,40){\oval(20,60)}\DashLine(250,40)(270,40){2}
\put(240,60){\circle{6}}\put(240,47){\circle{6}}\put(240,34){\circle{6}}
\put(240,21){\circle{6}}
\put(280,40){\oval(20,60)}
\put(280,60){\circle{6}}\put(280,47){\circle{6}}\put(280,34){\circle{6}}
\put(280,21){\circle{6}}
\DashLine(197,34)(185,34){2}\DashLine(185,34)(185,0){2}
\DashLine(200,18)(200,0){2}
\DashLine(203,21)(215,21){2}\DashLine(215,21)(215,0){2}
\DashLine(283,47)(295,47){2}\DashLine(295,47)(295,34){2}\DashLine(295,34)(283,34){2}
\DashLine(280,18)(280,5){2}\DashLine(280,5)(240,5){2}\DashLine(240,5)(240,18){2}

\end{picture}
\end{center}
This is not an algorithm, but it plays one on TV.

If I translate a diagram from linkbox form to likebox form, then translate back using the normal form
construction of the axioms, I don't get the diagram that I started with. I get something horrible instead.
This means that I should define an equivalence relation which will allow us to simplify likeboxes,
so that we can identify when two diagrams mean the same thing.

There are two different natural equivalences, both of which are useful. Two diagrams {\em have the same
kinematics} when the number and type of different complexes that they describe are the same. More
precisely, but still informally, the trees in two diagrams with the same kinematics are built from
corresponding elementary parts which can be attached in corresponding ways. This relation is useful
for enumeration algorithms, but it is too coarse an identification. Any non-recursive diagram, for example,
has the same kinematics as a finite collection of atoms, and any two diagrams with the same number of
complexes have the same kinematics.

The second notion of identity preserves likebox structure, and this is the one that I will use to
equate diagrams. Two diagrams are {\em equivalent} when they can be transformed into each other
through the following reversible moves:\\
\\
1. A likebox which contains itself $k$ times may be replaced by a likebox linked to itself $k$ times.\\
2. If a likebox contains precisely one other likebox, we may move an equivalence line from the inner to
the outer.\\
3. A  nonempty unlinked likebox may be deleted, its contents moved to its parents.\\
4. An empty likebox linked exactly once to likebox {\tt L} and possibly linked to $k$ other likeboxes
may be deleted, if all $k$ linkages are moved to {\tt L}.\\
5. If likeboxes ${\tt L_1 ... L_k}$ are a connected component of the equivalence line graph, we can
reconnect them in any way that preserves the number of distinct paths between any two likeboxes.\\
6. If ${\tt L_1 ... L_k}$ are a connected linkage component, we may permute their contents fixing the
equivalence lines, so long as the number of distinct linkage paths between any two destinations is the
same as the number of distinct paths from the sources.\\
\\
A {\em kinematic equivalence} is any iteration of these moves or their inverses. Rule 1 allows us to forbid
likeboxes from containing themselves, a good thing, since this is impossible to draw.  Rule 3 can be used
to remove unlinked likeboxes, and in conjunction with Rule 2, it allows us to remove likeboxes which
contain exactly one likebox. Rule 4 taken backwards allows us to attach a top-level atom to any likebox
by a line, or to break any equivalence line in two by inserting an atomic intermediate. Rules 5 and 6 let us
rearrange likeboxes and linkages in any connected component, which is an equivalnce so long as the distinct
paths between objects are unchanged.

To understand these transformations, it helps to study the kinematics in likebox form. The first thing
to do is to translate the notion of a link-path.\\
\\
A {\em path} in a likebox diagram is a non-backtracking sequence of likeboxes where successive ones are
either linked or one contains the other.\\
\\
Proposition: Every path in a likebox diagram determines a unique link-path in an axiomatic normal form
construction.\\
\\
Proof: In axiomatic normal form every likebox has an associated linkbox, so a path is a sequence
of linkboxes. If the hop is from likebox to it's container, the link-path passes through the state
attached to the container. If the path is along an equivalence, the link-path passes through the
two states that implement the equivalence, through the linkbox corresponding to the linkage. If the
path is from the container to one of the members, the link-path hops through the unique path from
the parent to the child. This map is one-to-one, since traversing any linkage, passing to a container,
or descending to a child each determines a unique path.\\
\\
Maximally branching link-paths are the same as maximally branching paths in a likebox diagram,
and these are the complexes. Since every state in an axiomatic normal form is linked to exactly one
other, axiomatic link-paths only branch to different switches of a linkbox. This means that branching
likebox paths only branch at linkage hops, not in containership hops. The branchings must still be
ordered lists, so that {\tt D<1-<2,1>,0>} is still distinct from {\tt D<2-<1,1>,0>} even when the
diagram is drawn with likeboxes.\\
\\
Each of the elementary equivalences allows us to locally redefine paths in a one-to-one way:\\
\\
1. Any path which hops from one of the $k$ interior copies to the exterior copy is replaced by a
path which travels along the $k$-th new equivalence line.\\
2. Any path which hops from the interior to the exterior is shortened by one hop.\\
3. No paths can stay in this likebox, so we can delete it and shorten the paths by one hop.\\
4. Any path which passes through the atom must leave by a linkage, so we can take the path
through {\tt L} instead, so long as we then leave through the corresponding linkage.\\
5. The path condition ensures that the paths are in correspondence.\\
6. ditto.\\
\\
So the kinematics of equivalent diagrams is identical, as we expect. It might seem that any two diagrams
with the same kinematics are equivalent, but it's not true. The diagram below has the same  kinematics
as twelve atoms, but is not equivalent to them.
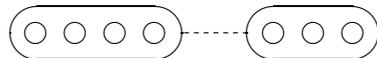
\begin{center}
\begin{picture}(240,30)(0,0)
\put(43,15){\oval(65,20)}\put(20,15){\circle{8}}\put(35,15){\circle{8}}\put(50,15){\circle{8}}\put(65,15){\circle{8}}
\DashLine(76,15)(100,15){2}
\put(125,15){\oval(50,20)}\put(110,15){\circle{8}}\put(125,15){\circle{8}}\put(140,15){\circle{8}}
\end{picture}
\end{center}
There is an equivalence invariant which shows this directly: the total number of unlinked empty
likeboxes doesn't change under any of the elementary moves.\\
\\
It is easy to see that permuting linkboxes, permuting switches in a linkbox, and permuting state-circles in
a switch produces equivalent diagrams, and this is why we can choose the order of each of these arbitrarily.\\
\\
Proposition: A likebox consisting of two atoms linked to otherwise unlinked top-level likeboxes is equivalent to
the union of the two likeboxes.\\
Proof: This is move 4 followed by move 2 on each of the two likeboxes.\\
Proposition: The normal form construction of likeboxes of likeboxes is equivalent to forming one likebox.\\ 
Proof: The last proposition performs the induction.\\
\\
Given a likebox form, we can reconstruct a reasonable size linkbox normal form equivalent to it, by the following
procedure:\\
1. Replace all likeboxes which contain themselves by likeboxes which are linked to themselves.\\
2. Replace all likeboxes which are linked to themselves by likeboxes linked through an atom.\\
3. Replace all likeboxes linked to each other $k$ times by likeboxes connected through $k-1$ atoms, and only
once directly.\\
4. Move all likeboxes to top-level by first linking them to a new top-level atom, and permuting their contents
with the atom.\\
\\
Now all likeboxes are either empty or only contain atoms. All linkages either join top-level likeboxes to
top-level likeboxes, atoms to top-level likeboxes, or atoms to atoms, and no linkage joins two things twice.\\
\\
5. Union any top-level likebox only joined to an atom with the likebox containing the atom. This collapses
the likeboxes maximally into chains.\\
6. If a likebox {\tt L} is joined to an atom {\tt A} and to something else too, insert an atom {\tt a} in
the linkage between {\tt A} and {\tt L}, replace the atom by a likebox containing an atom, and move the
linkage to the {\tt A} from the top-level likebox to the atom.\\ 
7. Whenever top-level likeboxes close a loop, break the loop by introducing two linked likeboxes along the
loop, each containing one atom. Then move the linkage from the outer likeboxes to the atoms they contain.\\
\\
After step 7, all nonempty likeboxes are top-level, and linked in linear chains. This implies that they
can be gathered into linkboxes, and this is a reasonable normal form for the diagram. Now if we build
linkboxes by the axioms, we see that we get a diagram equivalent to the linkboxes we want to build. I will
expand on this briefly in the next section.\\
\\
To convert mutual inhibition to likebox form, you can do this:
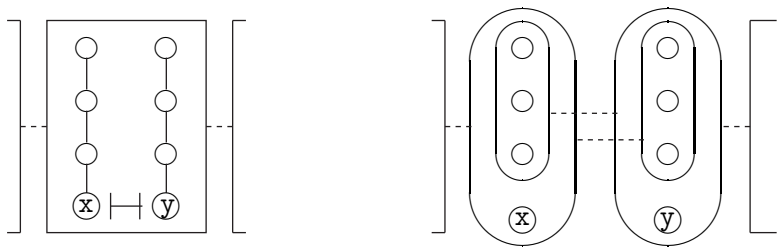
\begin{center}
\begin{picture}(300,110)
\Boxc(45,50)(60,80)
\put(30,80){\circle{8}}\Line(30,76)(30,64)\put(30,60){\circle{8}}\Line(30,56)(30,44)\put(30,40){\circle{8}}
\Line(30,36)(30,25)\put(30,20){\circle{10}}\put(27,18){\tt x}
\put(60,80){\circle{8}}\Line(60,76)(60,64)\put(60,60){\circle{8}}\Line(60,56)(60,44)\put(60,40){\circle{8}}
\Line(60,36)(60,25)\put(60,20){\circle{10}}\put(58,18){\tt y}
\Line(50,15)(50,25)\Line(51,20)(39,20)\Line(39,15)(39,25)
\DashLine(15,50)(5,50){2}\Line(5,10)(5,90)\Line(5,90)(0,90)\Line(5,10)(0,10)
\DashLine(75,50)(85,50){2}\Line(85,10)(85,90)\Line(85,90)(90,90)\Line(85,10)(90,10)

\put(195,60){\oval(20,60)}
\put(195,50){\oval(40,90)}
\put(195,80){\circle{8}}\put(195,60){\circle{8}}\put(195,40){\circle{8}}
\put(195,15){\circle{10}}\put(192,13){\tt x}
\put(250,60){\oval(20,60)}
\put(250,50){\oval(40,90)}
\put(250,80){\circle{8}}\put(250,60){\circle{8}}\put(250,40){\circle{8}}
\put(250,15){\circle{10}}\put(247,13){\tt y}
\DashLine(175,50)(165,50){2}\Line(165,10)(165,90)\Line(165,90)(160,90)\Line(160,10)(165,10)
\DashLine(270,50)(280,50){2}\Line(280,10)(280,90)\Line(280,10)(290,10)\Line(280,90)(290,90)
\DashLine(205,55)(230,55){2}
\DashLine(240,45)(215,45){2}

\end{picture}
\end{center}
But alas, this is not right. The linkages between the likeboxes close a loop, rendering this diagram
recursive. The constraint on this diagram is then that the complex formed from the loop must always
return to the same molecule. Implementing these constraints efficiently is the subject of the next
paper.

It is impossible to write a double inhibition in a normal form without duplicating likeboxes or
introducing spurious configurations. To write mutual inhibition, we need a diagram to attach the
switch composed of the likebox ${\tt S_1}$ and one extra state {\tt x} to the likebox ${\tt S_2}$
and one extra state {\tt y} without attaching {\tt x} to {\tt y}. Since we don't want to duplicate
the states in ${\tt S_1}$, we must have a path from ${\tt S_1}$ to {\tt y}. Similarly, we need a
path from ${\tt S_2}$ to ${\tt x}$ and from ${\tt S_1}$ to ${\tt S_2}$. These paths combine to
produce unwanted states, either connections between ${\tt x}$ and ${\tt S_1}$ or ${\tt y}$ and ${\tt S_2}$.
In the form we use above, both occur, since the example is recursive.

\section{Uniqueness}

The diagrams are formed by the axioms, and simultaneously include everything that can be formed from
them. This proves that the diagrams are the unique minimal realization of the axioms. In particular,
we cannot exclude all the crazy polymers.\\
\\
Theorem: (minimality) Any other kinematic formalism with constant size expressions for atoms, likeboxes,
and binary linkages can represent any normal form with an expression at most a constant factor larger
in size.\\
\\
Proof: I only need to build a diagram from the axioms, and check that the number of applications
is bounded by the size. The construction is the same as the heuristic one from before.

For every state in a diagram, we construct one atom for each circle representative, and link them.
This takes a number of steps proportional to the number of circles. Next form likeboxes from
each of the atoms whose circles are in the same switch, which takes as many axiom applications
as the sum of the log of the sizes of all the switches. As we have seen, this produces a diagram
equivalent to a likebox for every switch. Finally link the switches into linkboxes in a linear
chain. This last step produces a diagram equivalent to the original, and the number of operations
in the last step is equal to the total number of switches.\\
\\
So if we don't want polymers, we have to change the definition of linkage. The only natural alternative
is to make every self binding close a loop, so that binding something to itself does nothing. This
alternative formalism only describes objects with finitely many states, so it cannot describe a general
purpose computer, but it is a very nice way of describing finite state automata. The finite state version
of this formalism excludes dimerization as well as polymerization, and it is equivalent to D. Harel's
Higraphs\cite{Harel}.

\section{Appendix: Undecidability in Random Computation}

First I will review some classical results:\\
\\
Suppose program {\tt PREDICT(P)} purports to predict the eventual behavior of an arbitrary program {\tt P},
we prove that {\tt PREDICT} mispredicts at least one program.\\
\\
To do this, construct {\tt SPITE} to read it's own code and store it in a variable {\tt R}, calculate
{\tt PREDICT(R)}, and then do the opposite. {\tt SPITE} calls {\tt PREDICT} as a subroutine, so it is
slightly bigger than {\tt PREDICT} and takes a little longer to run.\\
\\
A proof following this schema is a proof by {\em free-will}. The only subtle part is that a program
can read its own code. This is true in any model of computation, but in the classical models, we have
to encode the program into it's input stream or into some variable. In any modern stored-instruction
computer there is nothing to do, since we can just look at the program's code in memory.\\
\\
Free-will is just a picturesque restatement of the method of Post, Godel, Church and Turing:\\
\\
If {\tt PREDICT} predicts whether a program halts, {\tt SPITE} calls predict on it's own source, and
if it is predicted to halt, {\tt SPITE} goes into an infinite loop, and halts otherwise. So the halting
problem is undecidable.\\
\\
If {\tt PREDICT} predicts whether {\tt SPITE} returns 1 or 0, {\tt SPITE} does the usual thing and
so the problem of determining which programs eventually return 1 and which return 0 is undecidable.
The set of all programs that return 1 is computably enumerable, as is the set of all programs which
return 0. So we have two computably enumerable sets which cannot be computably separated--- Kleene's
formulation of undecidability.\\
\\
If {\tt PREDICT} predicts whether {\tt SPITE} returns in polynomial time or exponential time, {\tt SPITE}
first ignores its input, and runs {\tt PREDICT} on its code. If {\tt PREDICT} says "polynomial", {\tt SPITE}
runs an exponential-time program on its input. If {\tt PREDICT} says "exponential", {\tt SPITE} just halts.
It is important to note that the running time of {\tt PREDICT} inside {\tt SPITE} doesn't depend on the size
of the input. So determining if a program runs in polynomial or exponential time is undecidable. This not a
classical result, but its one of many obvious applications of the classical method.\\
\\
If {\tt PREDICT} tells me whether an arbitrary theorem of Peano arithmetic is true or false, I write
{\tt SPITE} to deduce all truths of arithmetic using {\tt PREDICT}, and halt only if the arithmetic
statement that {\tt SPITE} never halts is true. This requires an embedding statements about a computer's
internal state into arithmetic, which can be done by a suitable encoding of bits into integers. So
determining whether an arbitrary statement in arithmetic is true or false is undecidable. If {\tt PREDICT}
deduces consequences of axioms, instead of truths, the same contradiction proves a strong version of Godel's
theorem.\\
\\
A simple consequence of classical free will is that a general computer running for time $t+C$ cannot be
predicted in time $t$ for large enough $t$. This implies that a system with accessible action angle variables
cannot compute, since transforming to action angle variables to accuracy $n$ digits by convergent power
series takes about $n$ steps, and allows us to time evolve the system to a given fixed accuracy up to
time $O(2^n)$.\\
\\
Now I will define stochastic computation.\\
\\
A {\em stochastic process} is an arbitrary sequence of bit-streams picked from a distribution $\rho$.
$\rho$ can be any probability distribution, even one with non-computable values.\\
\\
A {\em stochastic program} is a computer program which can get an arbitrary number of bits equally likely
to be zero or one. This process is an idealized random number generator, and it has a computable probability
distribution.\\
\\
A stochastic program {\tt P} is {\em certainly halting} if the probability that it runs forever is zero,
if for any $\epsilon$ there is an $N$ such that the probability of running past $N$ steps is less
than $\epsilon$.\\
\\
A program {\tt P} is {\em statistically indistinguishable} from {\tt S} if the distribution of outputs
of {\tt P} and {\tt S} is the same.\\
\\
To predict the results of a classical computation, we have to produce a definite answer. To predict a
random computation, I don't need to get the right answer every time. Only the distribution of answers
needs to match. This means that classical methods of proving undecidability don't work, and this is
the theorem below.\\
\\
Lemma: If {\tt SPITE} is a certainly halting program, its output distribution is a continuous function
of the output distribution of any stochastic process it contains as a subroutine.\\
\\
Proof: We choose $N$ large so that the probability of the program running more than $N$ steps is less
than $\epsilon/2$. {\tt SPITE} then only calls the subroutine at most $N$ times, so the distribution
of outputs only depends on the value of the first $N$ random variable answers. If we change the
distribution of each answer by a small enough amount, we can imagine that the first $N$ answers of the
new process are the same as the old with probability $1-\epsilon/2$, so we get the same overall output
with probability $1-\epsilon$.\\
\\
Theorem: Let {\tt SPITE} return a fixed bit-length answer, which stochastic process {\tt PREDICT}
tries to predict. {\tt SPITE} can call {\tt PREDICT} an arbitrary number of times with any input.
Nevertheless, there exists a {\tt PREDICT} whose output is statistically indistinguishable from
that of {\tt SPITE}, that is for which the prediction succeeds.\\
\\
Proof: The distribution of {\tt SPITE} is a continuous function of the distribution of {\tt PREDICT} by
the lemma. This defines a continuous function from the space of probability distributions on length $n$
bit strings to itself, and since this is a simplex of dimension $2^n-1$, the Brouwer fixed point theorem
guarantees that there is a distribution which is not moved at all. This fixed point is the {\tt PREDICT}
distribution we are looking for.\\
\\
So if {\tt SPITE} tries to spite {\tt PREDICT} just by running it, there is always a {\tt PREDICT} which
foils {\tt SPITE}! This means that to prove undecidability we have to look inside {\tt PREDICT} and see
what it is doing. Further, if we suppose {\tt PREDICT} can flip a coin which lands heads with probability
equal to an uncomputable real number that encodes the solution to the halting problem in its digits, it
can solve the halting problem to near certainty just by flipping enough coins. So we not only have to
look inside {\tt PREDICT}'s code, but we have to use the fact that the random numbers it uses have a
computable distribution. This is all we need, since we can now prove:\\
\\
Theorem:(probabilistic free will) Given any certainly halting stochastic program {\tt PREDICT} which
predicts the first bit of the eventual output of {\tt SPITE}, we can find a deterministic {\tt SPITE}
whose bit is mispredicted with probability at least $1/2 - \epsilon$ for any $\epsilon>0$.\\
\\
proof: We construct {\tt SPITE} to spite a deterministic version of {\tt PREDICT}, which uses a
fixed nonrandom string of bits in place of random numbers. We call this new {\tt PREDICT} with
every possible nonrandom string. From the outputs, we can construct an approximation to {\tt PREDICT}'s
distribution. The number of random bits used by {\tt PREDICT} is bounded for nearly all calls, so
we only need to search finitely many strings to get the distribution of answers to accuracy $\epsilon$.
Once we construct an approximation to the distribution of all answers, we return the least probable
value, either 0 or 1. This program is deterministic, so it returns the same answer every time, and
the best {\tt PREDICT} can do is get the answer right $1/2+\epsilon$ of the time.\\
\\
The noteworthy thing about this theorem is the abominable time bound. To spite a {\tt PREDICT} that
runs in $N$ steps on its source, {\tt SPITE} needs to run for $N2^N$ steps. It should be possible to
do much better, but I don't know how.

Despite that, this result proves stochastic undecidability. So we learn that the it is impossible to
use a random number generator to guess whether an arbitrary theorem of arithmetic is true or false
in a way that does significantly better than random chance. In biology, this means that probabilistic
methods will eventually fail at predicting the statistics of the behavior of a generic biological
system.

\end{document}